# A correct selection of asteroid families and confirmation of a nongravitational effect action

A. M. Kazantsev and L. V. Kazantseva

Astronomical Observatory of Kyiv Shevchenko National University, Ukraine
e-mail:  ankaz@observ.univ.kiev.ua

**Abstract**
The distribution of asteroid sizes by semimajor axes and the distribution of the number of asteroids by albedo values for separate families are used to selection the asteroid families more clearly. An analysis of the families identified by Masiero et al. (2013) was carried out with the use of the distributions of $D(a)$ and $N(p)$, there were defined correctly and incorrectly selected families. A decreasing of the mean albedo with semi-major axis increasing take place almost for all correctly selected families and which are not truncated by resonances. The decrease is statistically significant for the majority of these families. There is none family at a significant increase of albedo. This confirms our previous conclusion on existence of a non-gravitational effect in the asteroid belt which causes the spatial separation of asteroids with different albedos.

**Introduction**
Asteroid families were identified by K. Hirayama almost a century ago, but their investigation remains an important and interesting scientific problem by this time. The characteristics of different families help study the history of the asteroid belt formation, the physical properties of bodies located in various regions of the belt, etc. True identification of bodies belonging to different families is still an important problem. It is known that the asteroids in each family should have similar values of the proper orbital elements: semimajor axes $a'$, eccentricities $e'$, and inclinations $i'$. Just similarity of the proper elements serves as the basis for separating the members of different families from the background asteroids. The most well-known and reliable arrays of proper elements are contained in the papers authored by Milani and Knezevic and may be accessed at the NASA site (http//pds.nasa.gov).

Besides the simple selection of family members by on their proper elements, there are more refined approaches (e.g., the hierarchical clustering method (HCM), Zappala et al. 1990) to the identification of families. This method takes into account not only the similarity of the proper elements but the mutual spatial distances and velocities of asteroids as well.

Large data arrays containing albedos and sizes of asteroids have become available recently. These are IRAS-array which contains albedos and sizes for 2228 asteroids (IRAS, Tedersco et al. 2002), and WISE-array where data for more then 130 000 bodies are listed (WISE, Masiero et al. 2011). One can separate the asteroids of different families from the background more efficiently with the use of the asteroid albedos and sizes. For example, the authors of Masiero et al. (2013) for identification of asteroid families used HCM and the WISE data on asteroid albedos. The authors published the list of 76 families with a total of more than 38000 members. All these bodies are included in the WISE database.

**Selection criteria for asteroid families**

It is widely accepted that each individual family originates from the disruption of a single large body. At that the debris (family members) scatter in different directions, and their velocities vary from several tens to several hundreds of meters per second. The newly formed asteroid population fills the $a$–$e$–$i$ coordinate space in the region where the orbital elements of the parent body were located. The $a$, $e$, $i$ orbital elements of the newly formed family members should match or be very close to the corresponding present day proper elements. The more the number of debris, the higher must be their density with respect to the background bodies in the given coordinate space. Just the increased number density of bodies with respect to background asteroids was the first criterion for the families selection.

It is clear that the velocities of small debris are, on average, higher than the velocities of large ones. Therefore, the semimajor axes of orbits of small bodies will differ more from the semimajor axis of the parent body orbit. The orbital elements of the largest objects (object) may match almost coincide with the orbital elements of the parent body. This means that for any family the $D(a)$ distribution of sizes bodies by the semimajor axes should have a central maximum with descending wings on both side. The left wing (smaller $a$ values) may be slightly shorter than the right one (larger $a$ values). This is because the increase of $a$ is, as follows from the energy integral, somewhat weaker than the decrease one at isotropic dispersal with equal velocities. The initial $D(a)$ distribution may change somewhat under the influence of various nongravitational effects (NGEs). However, as any NGE primarily affects smaller bodies, the general form of the distribution (the central maximum with descending wings) is retained. Let us now formulate the first additional criterion: (a) for any family the $D(a)$ distribution of sizes bodies by the semimajor axes should have a central maximum with descending wings on both side. This is a necessary criterion. The right wing (larger $a$ values) is typically somewhat longer than the left one.

The *D*(*a*) distribution allows to determine more accurately the central body of the family (i.e., the asteroid after which the family is named). This asteroid should not only have the maximum size but also be located at the peak of the central maximum or near it. If the largest body of the family is located far from the maximum, it probably does not belong to this family. In that case one should choose the second largest asteroid located near the maximum. This asteroid will be the central body of the family.

Each parent body belongs to a certain taxonomic type. Therefore, the albedos of all members of the family would not greatly differ from the albedo typical for this type. This allows to formulate the second criterion: (b) for any family the *N*(*p*) distribution of the number of asteroids by their albedos should have a single primary maximum with a certain scatter on both side. This distribution may not have a pair (or more) of such maxima. The primary maximum may not be located at the very edge of the distribution. This is a necessary criterion.

Asteroid families may in principle form after collision of two asteroids of different taxonomic types. If the sizes of the collided bodies are comparable, the newly formed group of debris may have a bimodal *N*(*p*) distribution. However, this group should usually comprise two separate families. As the ranges of proper orbital elements *a*', *e*', *i*' of different families are rather narrow, even not very noticeable differences in the orbital elements of colliding asteroids result in different ranges of proper elements of debris belonging to different parent bodies. If a collision of two bodies with different taxonomic types and almost coinciding orbital elements happens, criterion (b) will allow to establish that fact.

It becomes possible to use the mentioned criteria (a) and (b) when large data arrays of asteroid albedos and sizes are available. In this paper we analyze the families listed in Masiero et al. (2013) with the use of the additional criteria based on asteroid albedos and sizes. At that Table 3 from Masiero et al. (2013) was used. The table was kindly rendered by Joseph Masiero, and we would like to thank him deeply for that.

**Analysis of the correctness of selection of asteroid families**
The authors of Masiero et al. (2013) publish a list of 76 families comprising a total of 38297 asteroids. All these bodies are included in the WISE database (Masiero et al. 2011). The names of the families in Masiero et al. (2013) correspond with the number of the largest asteroid given in the MPC catalogue. For example, the 004 family is known as Vesta family.

We are not going to analyze the procedure of asteroid family selection used in Masiero et al. (2013). We only note that the above-mentioned additional criteria were not applied by authors of Masiero et al. (2013). The analysis of Table 3 from Masiero et al. (2013) revealed a considerable number of errors in the asteroid family selection proceeding.

Based on the results of analysis of the selection, all the families may be divided into four groups. The first group consists of correctly identified families that are not visibly truncated by any resonances in the asteroid belt. The correctness of family selection one can see from the $D(a)$ and $N(p)$ distributions. For example the 163 and the 3811 families belong to the first group, and theirs $D(a)$ distributions are shown in Fig. 1. Those distributions clearly represent the family formation process described above. Of the 76 families listed in Masiero et al. (2013), 21 may be included into the first group.

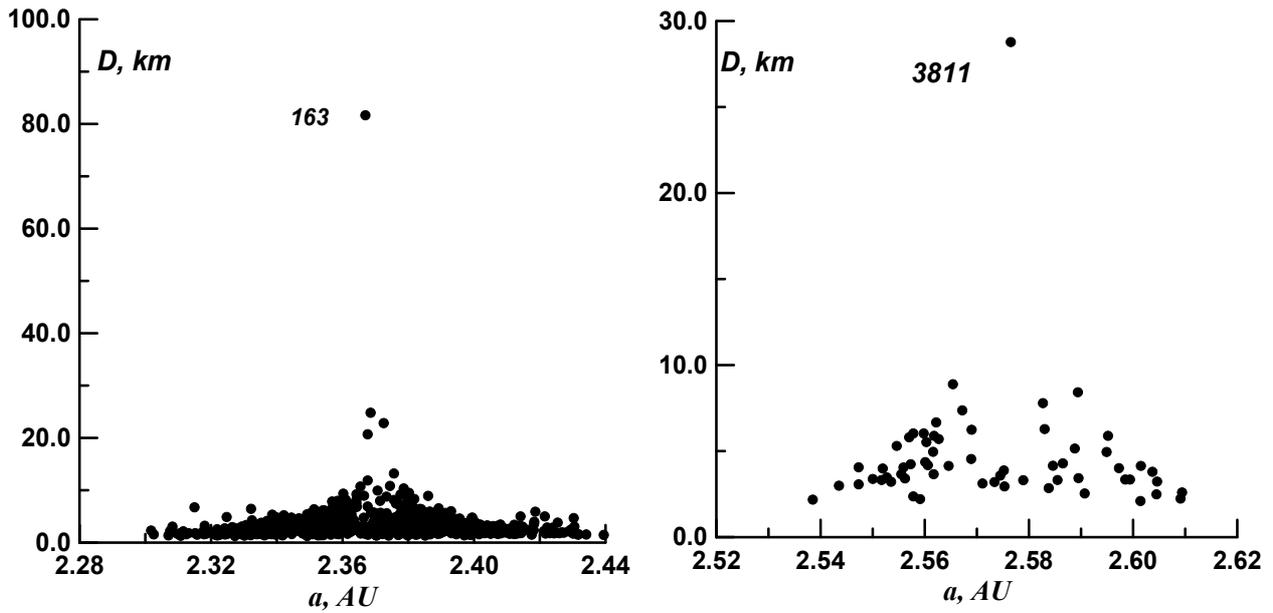

Fig. 1. The $D(a)$ distributions for two correctly identified families are not visibly truncated by resonances

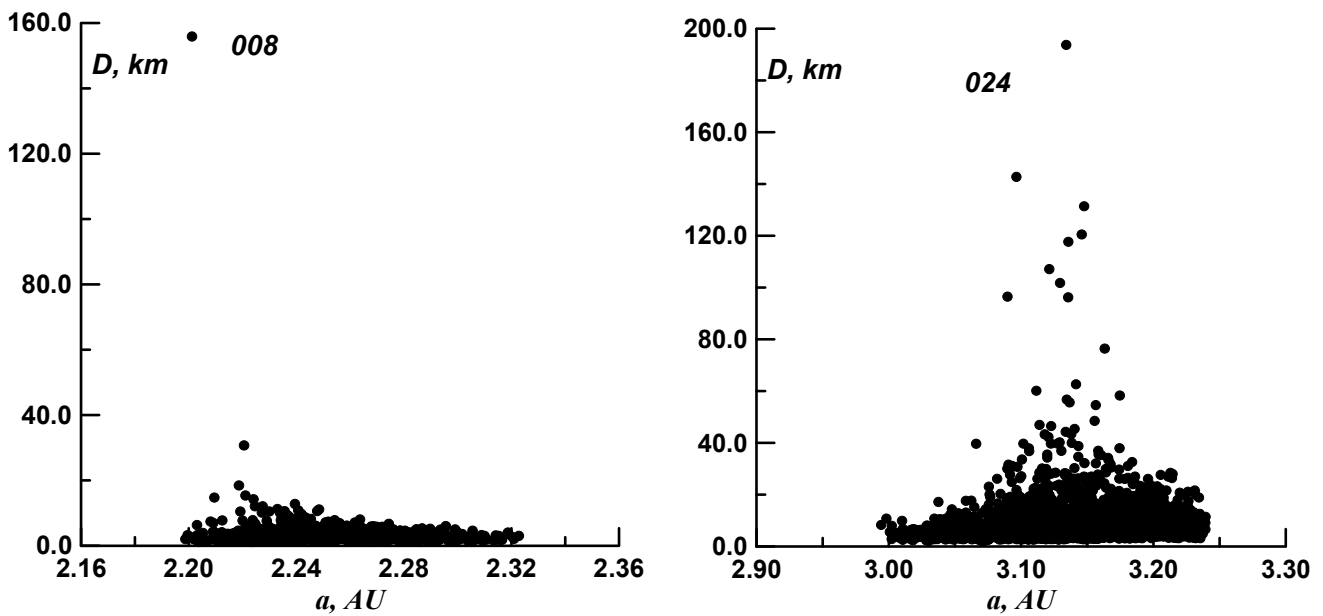

Fig. 2. The $D(a)$ distributions for two correctly identified families that are visibly truncated by resonances

The second group includes 23 correctly identified families that are visibly truncated under the influence of resonances. The 008 (Flora) family and 024 (Themis) family belong to this group. Theirs *D(a)* distribution are shown in Fig. 2. It is clearly seen that the 008 family has a central maximum, but the left wing is completely truncated by the $\nu_6$ resonance with Saturn. The right wing of the 024 family is truncated by resonance 2 : 1 with Jupiter about by 40%.

The third group consist of families that were selected with obvious errors. Several such families contain bodies that were included into other families from the list of 76 analyzed ones. Their *D(a)* distributions are similar to the classic ones (seen in Fig. 1 and in Fig. 2), but a significant number of irrelevant bodies ("interlopers") is easily seen. These "interlopers" may belong to the background asteroids or to other families. Two examples of this group are presented in Fig. 3. One can see that the 283 family really includes two families. The central maximum of the first family is located at *a* = 3.045 AU, of the second one – at *a* = 3.09 AU. The other distribution shows that semimajor axes for the 396 family should exceed 2.73 AU, and the bodies with *a* < 2.73 doesn't belong to the family.

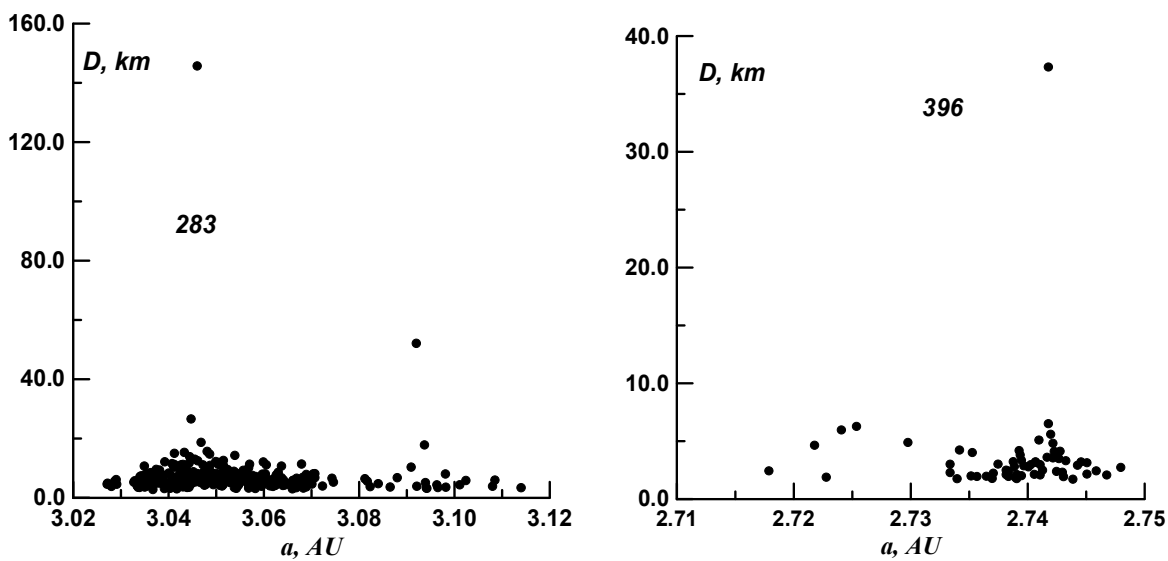

Fig. 3. The *D(a)* distributions for two families from the third group

Such bodies may sometimes be detected only with the use of the *N(p)* distribution that clearly reveals groups with different albedos. Figure 4 shows the *N(p)* distribution for the 2409 family. One can see the asteroids with albedos *p* ranging from 0.06 to 0.12 form a group that is isolated from the rest of the family. Of all the families listed in Masiero et al. (2013), 12 may be included into the third group.

The fourth group comprises the incorrectly identified families. Two examples of them are presented in Fig. 5. These are not numerous the 3567 family and the 4609 family. No central maximum and descending wings are observed. Such distributions may not be formed as a result of disruption of a single large body. Therefore, these groups of asteroids may not be considered as the families. The paper (Masiero et al. 2013) lists 20 families that belong to the fours group.

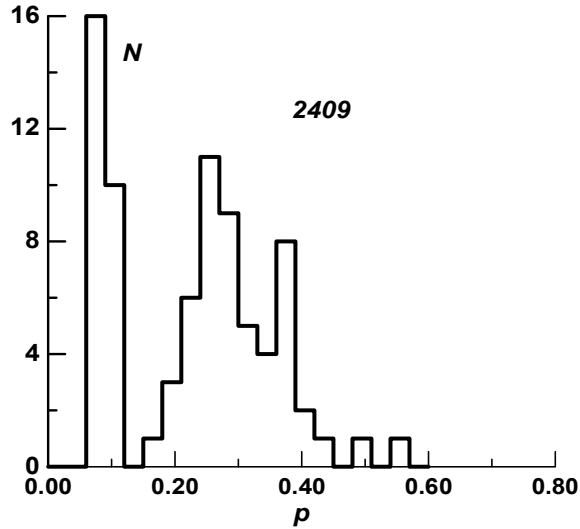

Fig. 4. The *N*(*p*) distribution for the 2409 family

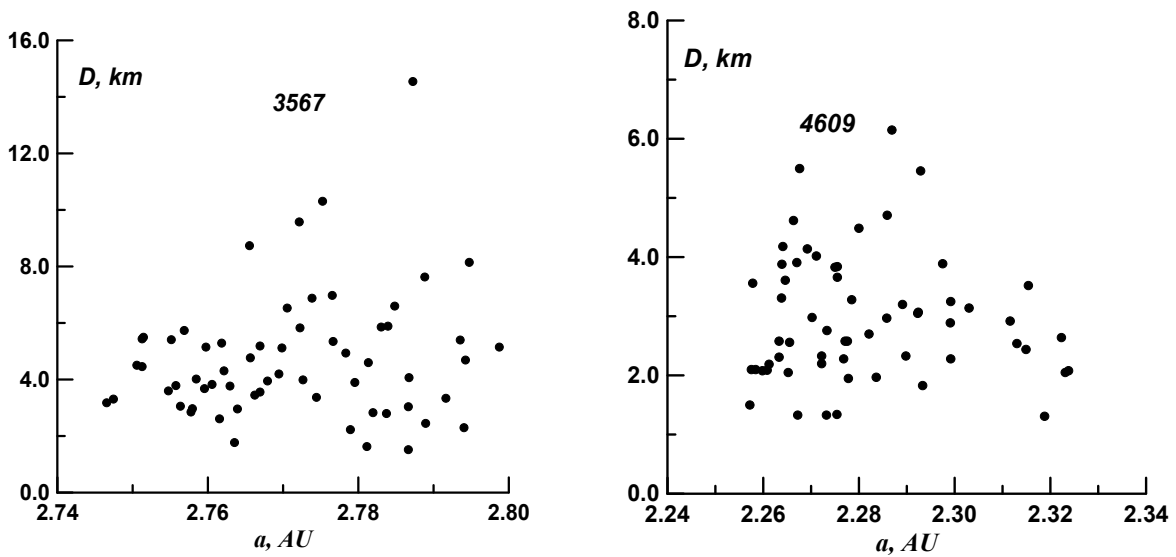

Fig. 5. The *D*(*a*) distributions for the incorrectly identified families

Figures 1 – 4 show that the *D*(*a*) and *N*(*p*) distributions allow one to confidently verify the correctness of asteroid family selection. The ranges of proper elements do not display the family structure, while these distributions make it clearly visible. Besides, the *D*(*a*) distribution allows one to unambiguously determine the primary asteroid in the family. Figure 6 shows the *D*(*a*) distribution for the 276 family.

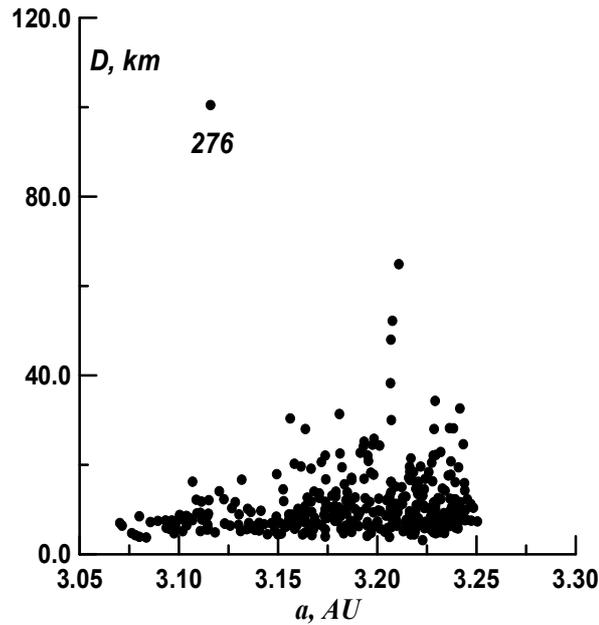

Fig. 6. The *D*(*a*) distributions for a family with incorrectly identified the primary asteroid

One can see the right wing of this family is truncated almost completely by the 2 : 1 resonance with Jupiter. The primary asteroid 276 is located far from the central maximum. Therefore, one can conclude that the asteroid 276 does not belong to this family. It is interesting to compare this distribution with the one for the 008 family (Fig. 2). Here the asteroid 008 is also located at the very inner edge of the range of semimajor axes, but it is nevertheless the central body of the family. It is obviously this fact may not be determined without the *D*(*a*) distribution.

**Manifestations of nongravitational effect in the families**

We have pointed out in our previous studies (Kazantsev. 2007, Kazantsev. 2008, Kazantsev and Kazantseva. 2008). that the influence of a certain NGE causes the spatial separation of asteroids with different albedos. The semimajor axes of orbits of asteroids with lower albedos are (on average) increasing with respect to the semimajor axes of orbits of bodies with higher albedos. This conclusion was based on the numerical calculations of the evolution of asteroid orbits and on the analysis of the *p*(*a*) dependences for separate asteroid families with using the IRAS data. The IRAS catalogue contains albedos and sizes of 2228 asteroids. Since this number is relatively small, the conclusion on the reality of the above-mentioned NGE should be confirmed with the use of a larger array. Such sample may be provided by the WISE catalogue that contains albedos and sizes of more than 130000 asteroids. Although the accuracy of the WISE catalogue is not very high, large data arrays allow one to obtain statistically significant results.

The influence of the above-mentioned NGE may be revealed by analyzing the dependences of the albedo on the semimajor axis for separate families. The $p(a)$ distribution of debris right after the parent body disruption should be close to a uniform one. Gravitational perturbations during the process of evolution of orbits may also not affect this distribution qualitatively. Thus, an evident increasing or decreasing of average albedo along semimajor axis for one or another family may be only considered as a result of a NGE action.

It is obvious that only the $p(a)$ dependences for correctly identified families that are not visibly truncated by resonances (i.e., families from the first group) should be considered for the revealing of the NGE. A family with a noticeable truncated wing (wings) has not only an incomplete range of the proper semimajor axes, but it may have an imperfect range of albedos as well. It is very difficult to determine the fact of a NGE action (or its absence) from the $p(a)$ dependence for such family.

The averaged linear $p(a)$ dependences were plotted for all families of the first group:

$$p(a) = b_1 \times p + b_0 \qquad (1)$$

A probable influence of the NGE on the asteroids of a certain family is detected by the sign and significance of coefficient $b_1$. If this coefficient is negative and has a sufficiently high statistical significance, one may conclude that the NGE influence on the bodies of the family is actually manifested. The $2\sigma$ level (corresponds to a quantile of 0.05) is defined as sufficient in mathematical statistics. In other words, if a certain parameter is significant at the $2\sigma$ level, the probability that its value is random does not exceed 5%.

Certain properties of the families belonging to the first group (specifically, the central asteroid number; the quantity of bodies in the family $N$; the ranges of proper elements and albedo $p$; the values of coefficients $b_1$ of the averaged $p(a)$ dependence; and the significance level of these coefficients $k\sigma$) are listed in the table. It should be noted that the dependences like (1) were plotted for the albedo range of 0.02–0.60. The albedo values falling out of this range may legitimately be considered erroneous.

The table shows that the $b_1$ coefficients are positive for six families, but their statistical singnificances are not rather high. Besides, five $b_1$ values of these six are smaller than 0.1 (i.e., the $p(a)$ distribution is close to the uniform one). The $b1$ coefficients are negative for the remaining 15 families, and eight of them have their significance at or above the $2\sigma$ level. The significance level is especially high in the case of the 221 (Eos) family. The $p(a)$ distributions and the averaged dependences (dashed lines) for the 928 family and for the 3985 family are shown in Fig. 7.

Table. Parameters for the correctly identified families that are not visibly truncated by any resonances

| Name | N | $\Delta a'$, AU | $\Delta e'$ | $\Delta i'$ (гр.) | $\Delta p$ | $b_1$ | $k\sigma$ |
|---|---|---|---|---|---|---|---|
| 004 | 1331 | 2.258-2.462 | 0.081-0.124 | 5.64-7.56 | 0.067-0.843 | –0.28 | 4.0 |
| 020 | 203 | 2.369-2.448 | 0.147-0.174 | 1.22-1.80 | 0.072-0.581 | –0.05 | 0.2 |
| 135 | 740 | 2.320-2.476 | 0.160-0.219 | 2.00-2.50 | 0.123-0.548 | –0.33 | 3.5 |
| 1646 | 46 | 2.316-2.370 | 0.093-0.104 | 7.70-8.45 | 0.067-0.423 | –0.46 | 0.6 |
| 012 | 269 | 2.295-2.448 | 0.174-0.212 | 8.36-10.96 | 0.029-0.152 | +0.02 | 0.5 |
| 1715 | 178 | 2.354-2.453 | 0.204-0.252 | 10.07-11.91 | 0.029-0.154 | +0.07 | 0.9 |
| 163 | 1093 | 2.302-2.440 | 0.189-0.233 | 4.12-6.42 | 0.021-0.151 | –0.03 | 0.2 |
| 005 | 94 | 2.563-2.604 | 0.182-0.217 | 3.98-4.90 | 0.066-0.629 | –0.35 | 3.4 |
| 808 | 90 | 2.711-2.787 | 0.127-0.139 | 4.76-5.34 | 0.071-0.543 | +0.27 | 0.7 |
| 363 | 512 | 2.689-2.789 | 0.025-0.060 | 4.77-5.73 | 0.024-0.149 | +0.06 | 1.5 |
| 272 | 861 | 2.748-2.819 | 0.040-0.060 | 3.90-5.18 | 0.01-0.153 | –0.01 | 0.3 |
| 144 | 184 | 2.614-2.709 | 0.172-0.200 | 3.39-4.50 | 0.026-0.129 | –0.01 | 0.2 |
| 3811 | 61 | 2.539-2.610 | 0.103-0.111 | 10.67-10.95 | 0.019-0.131 | –0.22 | 2.0 |
| 569 | 357 | 2.594-2.704 | 0.170-0.191 | 1.75-2.85 | 0.024-0.123 | –0.01 | 0.2 |
| 1128 | 201 | 2.754-2.816 | 0.045-0.052 | 0.51-1.13 | 0.023-0.102 | +0.01 | 0.2 |
| 221 | 5718 | 2.937-3.158 | 0.022-0.144 | 8.50-12.82 | 0.066-0.558 | –0.18 | 13.4 |
| 3985 | 126 | 2.835-2.880 | 0.119-0.126 | 14.83-15.24 | 0.070-0.335 | –1.49 | 2.8 |
| 3330 | 734 | 3.100-3.189 | 0.179-0.222 | 9.39-11.02 | 0.014-0.143 | +0.03 | 0.7 |
| 283 | 315 | 3.027-3.072 | 0.102-0.130 | 8.69-9.65 | 0.016-0.121 | –0.20 | 2.6 |
| 702 | 49 | 3.158-3.247 | 0.006-0.026 | 20.88-22.42 | 0.029-0.146 | –0.22 | 0.8 |
| 928 | 85 | 3.091-3.184 | 0.181-0.206 | 15.88-17.03 | 0.017-0.110 | –0.22 | 2.7 |

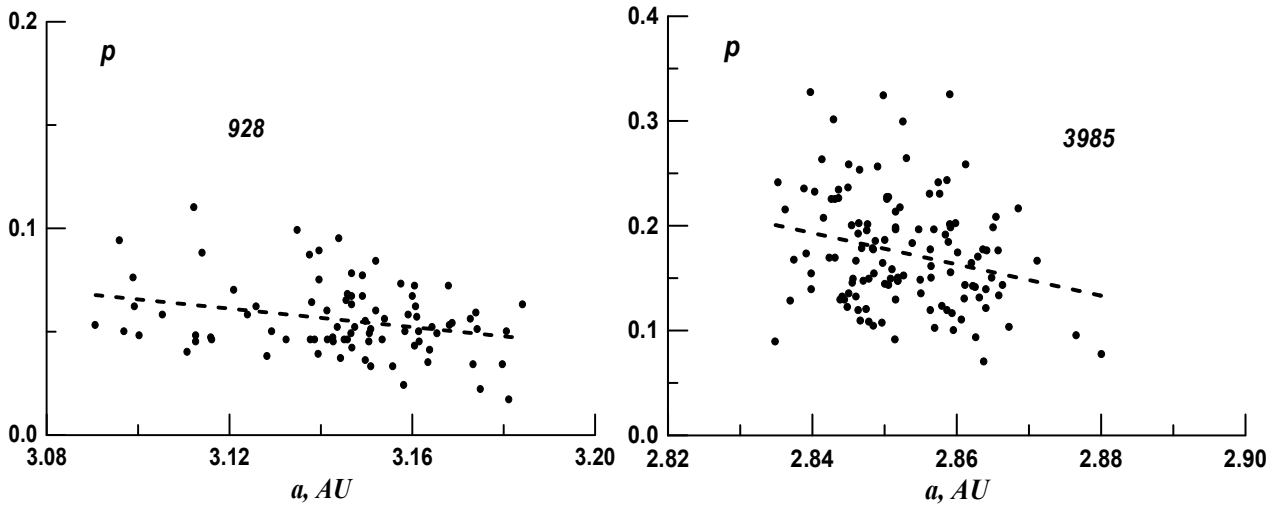

Fig. 7. The $p(a)$ distributions and the averaged dependences (dashed lines) for two families

It should mention especially the 3985 family with the steepest reduction in albedo. Although this family is relatively small, it has a rather high statistical validity of the $b_1$ coefficient. This family is located in the region of values of proper elements $a'$–$e'$–$i'$ with no other neighboring families and a background asteroid density is tens of times lower than the spatial density of bodies belonging to the family. Therefore, the probability of "interlopers" entering the 3985 family is extremely small. Thus, the data presented in the table confirm existence of a non-gravitational effect in the asteroid belt which causes the spatial separation of bodies with different albedos. A similar study in a little shorter form is presented in Kazantsev and Kazantseva. (2014).

It is clear, a physical mechanism of the NGE should be found. A short qualitative description of a possible mechanism is pointed in Kazantsev. (2008). More detail study of such problem is necessary.

**Conclusions**

The proposed additional criteria for identifying the asteroid families based on the $D(a)$ and $N(p)$ distributions prove to be rather efficient in correctly screening out the foreign bodies, verifying the completeness of the family, and determining its primary asteroid. If these criteria are not used, a significant proportion of families may be isolated with noticeable errors.

A reduction in the mean albedo with increasing semimajor axis is observed for almost all correctly identified families that are not truncated by resonances. This reduction is statistically significant for the majority of these families. Not a single family exhibits a statistically significant increase in albedo. This confirms our previous conclusions of existence of a non-gravitational effect in the asteroid belt which causes the spatial separation of bodies with different albedos.